\PassOptionsToPackage{dvipsnames, table}{xcolor}
\documentclass[10pt,twocolumn,letterpaper]{article}

\usepackage{cvpr}              

\usepackage[table,xcdraw]{xcolor}
\definecolor{cvprblue}{rgb}{0.21,0.49,0.74}
\usepackage[pagebackref,breaklinks,colorlinks,allcolors=cvprblue]{hyperref}
\usepackage{multirow,multicol,makecell,hhline,booktabs} 
\usepackage{array}
\usepackage[linesnumbered,ruled]{algorithm2e}

\newcommand{\Bmat}{{\bf B}}

\newcommand{\Emat}[0]{{{\bf E}}}
\newcommand{\Fmat}[0]{{{\bf F}}}

\newcommand{\Hmat}[0]{{{\bf H}}}
\newcommand{\Imat}{{\bf I}}

\newcommand{\Kmat}[0]{{{\bf K}}}

\newcommand{\Pmat}[0]{{{\bf P}}}
\newcommand{\Qmat}[0]{{{\bf Q}}}

\newcommand{\Umat}{{{\bf U}}}
\newcommand{\Vmat}[0]{{{\bf V}}}
\newcommand{\Wmat}[0]{{{\bf W}}}
\newcommand{\Xmat}{{\bf X}}
\newcommand{\Ymat}[0]{{{\bf Y}}}
\newcommand{\Zmat}{{\bf Z}}

\newcommand{\xv}{\boldsymbol{x}}
\newcommand{\yv}{\boldsymbol{y}}

\newcommand{\zv}{\boldsymbol{z}}

\newcommand{\Phimat}{\boldsymbol{\Phi}}

\newcommand{\epsilonv}{\boldsymbol{\epsilon}}

\newcommand{\tsp}{^{\top}}

\usepackage{amsmath}
\makeatletter
\let\save@mathaccent\mathaccent
\newcommand*\if@single[3]{%
  \setbox0\hbox{${\mathaccent"0362{#1}}^H$}%
  \setbox2\hbox{${\mathaccent"0362{\kern0pt#1}}^H$}%
  \ifdim\ht0=\ht2 #3\else #2\fi
  }
\newcommand*\rel@kern[1]{\kern#1\dimexpr\macc@kerna}
\newcommand*\widebar[1]{\@ifnextchar^{{\wide@bar{#1}{0}}}{\wide@bar{#1}{1}}}
\newcommand*\wide@bar[2]{\if@single{#1}{\wide@bar@{#1}{#2}{1}}{\wide@bar@{#1}{#2}{2}}}
\newcommand*\wide@bar@[3]{%
  \begingroup
  \def\mathaccent##1##2{%
    \let\mathaccent\save@mathaccent
    \if#32 \let\macc@nucleus\first@char \fi
    \setbox\z@\hbox{$\macc@style{\macc@nucleus}_{}$}%
    \setbox\tw@\hbox{$\macc@style{\macc@nucleus}{}_{}$}%
    \dimen@\wd\tw@
    \advance\dimen@-\wd\z@
    \divide\dimen@ 3
    \@tempdima\wd\tw@
    \advance\@tempdima-\scriptspace
    \divide\@tempdima 10
    \advance\dimen@-\@tempdima
    \ifdim\dimen@>\z@ \dimen@0pt\fi
    \rel@kern{0.6}\kern-\dimen@
    \if#31
      \overline{\rel@kern{-0.6}\kern\dimen@\macc@nucleus\rel@kern{0.4}\kern\dimen@}%
      \advance\dimen@0.4\dimexpr\macc@kerna
      \let\final@kern#2%
      \ifdim\dimen@<\z@ \let\final@kern1\fi
      \if\final@kern1 \kern-\dimen@\fi
    \else
      \overline{\rel@kern{-0.6}\kern\dimen@#1}%
    \fi
  }%
  \macc@depth\@ne
  \let\math@bgroup\@empty \let\math@egroup\macc@set@skewchar
  \mathsurround\z@ \frozen@everymath{\mathgroup\macc@group\relax}%
  \macc@set@skewchar\relax
  \let\mathaccentV\macc@nested@a
  \if#31
    \macc@nested@a\relax111{#1}%
  \else
    \def\gobble@till@marker##1\endmarker{}%
    \futurelet\first@char\gobble@till@marker#1\endmarker
    \ifcat\noexpand\first@char A\else
      \def\first@char{}%
    \fi
    \macc@nested@a\relax111{\first@char}%
  \fi
  \endgroup
}
\makeatother

\def\paperID{12380} 
\def\confName{CVPR}
\def\confYear{2025}

\title{Proximal Algorithm Unrolling:\\
Flexible and Efficient Reconstruction Networks for Single-Pixel Imaging}

\author{
Ping Wang$^{1,2}$\thanks{Equal contribution.} ~\enspace Lishun Wang$^{1}$\footnotemark[1] ~\enspace Gang Qu$^{1}$ \enspace Xiaodong Wang$^{1,2}$ \enspace Yulun Zhang$^{3}$ \enspace Xin Yuan$^{1}$\thanks{Corresponding author: Xin Yuan, xyuan@westlake.edu.cn}\\
$^{1}$Westlake University~~$^{2}$Zhejiang University~~$^{3}$Shanghai Jiao Tong University
\\
}

\begin{document}
\maketitle
\begin{abstract}
Deep-unrolling and plug-and-play (PnP) approaches have become the de-facto standard solvers for single-pixel imaging (SPI) inverse problem. 
PnP approaches, a class of iterative algorithms where regularization is implicitly performed by an off-the-shelf deep denoiser, are flexible for varying compression ratios (CRs) but are limited in reconstruction accuracy and speed. 
Conversely, unrolling approaches, a class of multi-stage neural networks where a truncated iterative optimization process is transformed into an end-to-end trainable network, typically achieve better accuracy with faster inference but require fine-tuning or even retraining when CR changes.
In this paper, we address the challenge of integrating the strengths of both classes of solvers.
To this end, we design an efficient deep image restorer (DIR) for the unrolling of HQS (half quadratic splitting) and ADMM (alternating direction method of multipliers).
More importantly, a general proximal trajectory (PT) loss function is proposed to train HQS/ADMM-unrolling networks such that learned DIR approximates the proximal operator of an ideal explicit restoration regularizer.
Extensive experiments demonstrate that, the resulting proximal unrolling networks can not only flexibly handle varying CRs with a single model like PnP algorithms, but also outperform previous CR-specific unrolling networks in both reconstruction accuracy and speed.
Source codes and models are available at \url{https://github.com/pwangcs/ProxUnroll}.

\end{abstract}    
\section{Introduction}
\label{sec:intro}
How many pixels does your camera actually need? Modern cameras typically require as many pixels as the image resolution demands. By contrast, single-pixel cameras need only one pixel, a light-sensitive detector, to capture images through multiple measurements.
Single-pixel imaging (SPI)~\cite{duarte2008single,watts2014terahertz,zhang2017hadamard,edgar2019principles,hahamovich2021single} technique measures the total light intensity of a scene coded by a series of spatially resolved masks on a single-pixel detector.
Known masks and measurements, one can recover an image with the same resolution as the masks.
SPI offers a competitive edge over mainstream CCD and CMOS cameras in certain scenarios, \eg, terahertz imaging~\cite{watts2014terahertz} and non-visible 3D imaging~\cite{sun20133d}.

\begin{figure}[t]
\centering
\includegraphics[width=1\linewidth]{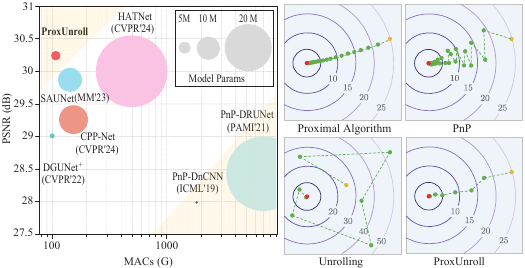}
\vspace{-4mm}
\caption{Average PSNR (left) of different methods at compressive ratios (CRs) $\{0.01,0.04,0.10,0.25\}$ in which only methods in the yellow area can flexibly handle varying CRs with a single model. Optimization trajectories (right) of four classes of methods for solving inverse problems.
Our ProxUnroll efficiently achieves SOTA performance with high flexibility and fast convergence.}
\vspace{-4mm}
\label{fig:summary}
\end{figure}
\begin{figure}[t]
\centering
\vspace{-4mm}
\includegraphics[width=\linewidth]{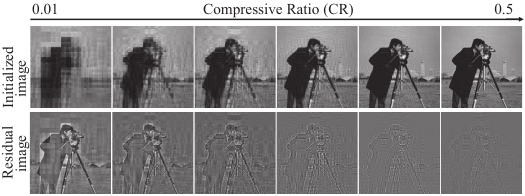}
\vspace{-5mm}
\caption{Degradations under different compressive ratios (CRs).}
\vspace{-6mm}
\label{fig:degradation}
\end{figure}

SPI cameras typical work under sub-Nyquist sampling rate to save acquisition time, leading to a compressed sensing problem~\cite{candes2006robust,4016283} of reconstructing a clean image $\xv \!\in\!\mathbb{R}^n$ from its few measurements $\yv \!\in\!\mathbb{R}^m$ obtained by the imaging model $\yv \!=\! {\Phimat} {\xv} \!+\! \epsilonv $, where $\Phimat \!\in\!\mathbb{R}^{m \times n}$ $(m \!\ll\! n)$ is the measurement (\ie, degradation) matrix, $\epsilonv \!\in\!\mathbb{R}^m$ the measurement noise.
The inverse problem is formulated as
\begin{equation}
\label{eq:prob}
\setlength{\abovedisplayskip}{0.1cm}
\setlength{\belowdisplayskip}{0.1cm}
\textstyle \hat{\xv} = \mathop{\arg\min}\limits_{\xv}f(\xv)+ \lambda g(\xv), 
\end{equation}
where $f(x)\!=\!\frac{1}{2} \Vert {\yv} \!-\! {\Phimat} {\xv} \Vert_{2}^{2}$ is a data fidelity term, $\lambda g(x)$ a scaled regularization term that represents a prior knowledge on $\xv$, \eg, sparsity~\cite{figueiredo2007gradient}.
In SPI cameras, $\Phimat$ is usually set to be row-orthogonal and each row represents an optical mask.
As a adjustable parameter, compressive ratio (CR) $\eta \!=\!{m \mathord{\left/
 {\vphantom {m n}} \right.
 \kern-\nulldelimiterspace} n}$ affects the ill-posedness of \cref{eq:prob}, namely the degree to which images are degraded as illustrated in~\cref{fig:degradation}.

A fruitful solution to \cref{eq:prob} is first-order proximal algorithms~\cite{combettes2011proximal}, \eg, half quadratic splitting (HQS)~\cite{geman1995nonlinear} method and alternating direction method of multipliers (ADMM)~\cite{Metzler2016FromDT}, which individually perform the proximal operator ${\tt Prox}$ on $f$ and $ g$. ${\tt Prox}_{h}$ is defined as 
\begin{equation}
\label{eq:prox}
\setlength{\abovedisplayskip}{0.1cm}
\setlength{\belowdisplayskip}{0.1cm}
{\tt Prox}_{h}(\zv) =  \mathop{\arg\min}\limits_{\xv} h(\xv) +\frac{1}{2} \Vert {\xv} \!-\! {\zv} \Vert_{2}^{2}.
\end{equation}
The properties of proximal operator guarantee the convergence of iterative algorithms~\cite{parikh2014proximal}.
${\tt Prox}_{f}$ has a closed form for a specific algorithm.
${\tt Prox}_{g}$ can be loosely interpreted as a denoising step, which is explicitly implemented by the regularizer ${g}$ or implicitly by a deep neural network.
By integrating proximal algorithms and deep neural networks, deep-unrolling (or unfolding)~\cite{zhang2018ista,aggarwal2018modl,dong2018denoising,bertocchi2020deep,monga2021algorithm} and plug-and-play (PnP)~\cite{zhang2017learning,sun2019block,sun2021scalable,zhang2021plug,10004791} approaches have become the de-facto standard solvers for imaging inverse problems.
PnP approaches constitute a class of iterative algorithms where regularization is implicitly performed by an off-the-shelf deep Gaussian denoiser~\cite{zhang2017beyond,zhang2018ffdnet,zhang2021plug}. 
Unrolling approaches constitute a class of multi-stage neural networks where a truncated iterative optimization process is transformed into an end-to-end trainable network.

In SPI cameras, degradations are signal-dependent and CR-dependent as illustrated in~\cref{fig:degradation}.
PnP algorithms~\cite{tian2022plug} have flexibility for varying CRs but typically achieve moderate accuracy with slow inference and cumbersome parameter adjustment.
Latest works~\cite{ryu2019plug,hurault2022gradient,hurault2022proximal,fangs,hu2024stochastic} reveal that, PnP deep denoiser or even deep restorer can approximate a proximal operator under certain assumptions or constraints to guarantee the convergence for various imaging tasks and varying degradations. 
Conversely, unrolling networks~\cite{wang2023saunet,qu2024dual} are typically trained end-to-end for a specific CR to achieve state-of-the-art (SOTA) accuracy with fast inference.
However, when CR changes, they must be fine-tuned or even retrained due to the out-of-distribution (OOD) problem, thereby lacking flexibility.
Besides, what has been learned from data is not interpretable for unrolling networks. 
As summarized in~\cref{tab:method}, PnP algorithms are highly flexible but are limited in accuracy and speed, whereas unrolling networks are too specialized to flexibly handle varying CRs.   
What hinders the flexibility and interpretability of unrolling networks is an open question.


In this paper, we address the challenge of integrating the strengths of PnP and unrolling approaches.
The flexibility and interpretability are originated from proximal algorithms.
{\em The flexibility and interpretability of PnP and unrolling approaches depend on the extent to which the used neural networks approach proximal operators}.
To this end, we propose a proximal unrolling (ProxUnroll) approach to endow unrolling networks with flexibility and interpretability.
Main contributions are summarized as follows:
\begin{itemize}
\item We introduce HQS- and ADMM-unrolling networks for SPI, composed of a few iterations of an explicit proximal operator and an implicit deep image restorer (DIR).

\item We design an efficient CNN-Transformer hybrid architecture as DIR, where the integration of window attention and dynamic CNN captures both low- and high-frequency information with high input dependency, and channel attention functions as a memory mechanism to propagate informative representations between adjacent iterations.

\item More importantly, we propose a generic proximal trajectory (PT) loss function to train HQS/ADMM-unrolling networks into HQS/ADMM-ProxUnroll networks where learned DIRs approximate the proximal operator of an ideal explicit restoration regularizer.

\item Extensive experiments demonstrate that the proposed HQS/ADMM-ProxUnroll efficiently achieves not only SOTA performance and fast convergence, but also the same flexibility as PnP algorithms (see~\cref{fig:summary}).
\end{itemize}

\begin{table}[t]
\vspace{-4mm}
\caption{Comprehensive comparison among proximal algorithms, PnP algorithms, unrolling networks, and our ProxUnroll.} 
\vspace{-2mm}
\centering
\label{tab:method}
\setlength{\tabcolsep}{3pt}
\scalebox{0.85}{
\begin{tabular}{l|llll}
\toprule[0.15em]
Method & Accuracy & Speed & Flexibility & Interpretability  \\ 
\hline \hline
Proximal  &   Low & Low  &  High  &  High   \\
PnP     &  Middle  & Middle &  High  &  Middle \\ 
Unrolling &   High & High   &  Low  & Low \\ 
ProxUnroll  & High  & High  & High & High \\  
\bottomrule[0.15em]
\end{tabular}
}
\vspace{-4mm}
\end{table}

\section{Related Works}
\label{sec:related}
{\bf SPI Reconstruction.}
SPI reconstruction is a classical inverse problem in the field of compressive imaging~\cite{RN5318,wang2023full,wang2024hierarchical,wang2023deep}. Mainstream solvers involve iterative optimization algorithms~\cite{figueiredo2007gradient,4587391,he2009exploiting,blumensath2009iterative,beck2009fast,kim2010compressed,yang2011alternating,dong2014compressive,Metzler2016FromDT}, PnP algorithms~\cite{zhang2021plug,hurault2022gradient,hurault2022proximal,fangs,hu2024stochastic}, single-stage neural networks~\cite{kulkarni2016reconnet,shi2019image,shi2019scalable,yao2019dr2}, and unrolling (multi-stage) neural networks~\cite{metzler2017learned,zhang2018ista,yang2018admm,zhang2020optimization,zhang2020amp,shen2022transcs,song2021memory,you2021coast,mou2022deep,ye2023csformer,song2023optimization,wang2023saunet,wang2024ufc,guo2024cpp,qu2024dual}.
Iterative algorithms employ hand-crafted regularizers, \eg, sparsity~\cite{figueiredo2007gradient}, total variation~\cite{4587391}, non-local low rank~\cite{dong2014compressive,yuan2016generalized}, with a proximal algorithm, \eg, iterative shrinkage thresholding algorithm (ISTA)~\cite{beck2009fast}, approximate message passing (AMP)~\cite{Metzler2016FromDT}, HQS~\cite{geman1995nonlinear}, and ADMM~\cite{yang2011alternating}. 
Single-stage networks generally achieve inferior performance due to the insufficient utilization of imaging model information.
PnP algorithms are flexible for varying CRs and unrolling networks achieve superior performance, thus making both the de-facto standard tools for SPI reconstruction. 

\noindent{\bf Proximal learning.}
The objective of proximal learning is to train a neural network as the proximal operator of a explicit regularization function.
Regularization by denoising~\cite{romano2017little} shows
that under homogeneity, nonexpansiveness and Jacobian
symmetry conditions, a denoiser can be written as a gradient descent step on a convex function.
However, such conditions are unrealistic for deep denoisers.
Recently, a new type of gradient denoisers~\cite{hurault2022gradient,hurault2022proximal,fangs} has been proposed by training a denoiser as an explicit gradient step on a functional parameterized by a deep neural network.
However, these denoisers must either be a contractive gradient~\cite{hurault2022gradient,hurault2022proximal} or be constrained to input convex neural networks (ICNN)~\cite{fangs}, inevitably sacrificing the expressivity.
Proximal learning without assumptions and constraints remains an open challenge.

\section{Preliminaries}
\label{sec:preliminaries}
In this section, we first introduce the inverse problem of SPI. Then, we extend two representative proximal algorithms (\ie, HQS and ADMM) to solve this problem. Finally, we compare PnP approaches with unrolling approaches.

\subsection{SPI Inverse Problem}
\label{subsec:spi}
Typical SPI cameras capture 2D images. Following \cref{eq:prob,eq:prox}, processing a 2D image in its vectorized form leads to large matrix calculation problems for all $\Phimat$-related operations, particularly during model training.
To this end, most previous works~\cite{zhang2018ista,zhang2020optimization,zhang2020amp,song2021memory,mou2022deep,ye2023csformer,song2023optimization,wang2024ufc,guo2024cpp,you2021coast} split a digital image into non-overlapping blocks to reduce the vectorized size, but block-wise sampling is impractical for real SPI cameras. 
A practical solution is Kronecker compressed sensing~\cite{duarte20212kronecker,caiafa2013multi}, which processes 2D images in matrix form rather than vector form.
Mathematically, $\yv \!=\! \Phimat \xv \!+\! \epsilonv$ is equivalent to $\Ymat \!\!=\!\! \Hmat \Xmat {\Wmat\tsp\!} \!+\! \Emat$ with $\yv\!\!=\!\!\mbox{vec}(\Ymat), \xv\!\!=\!\!\mbox{vec}(\Xmat), \epsilonv\!\!=\!\!\mbox{vec}(\Emat), \Phimat \!\!=\!\! \Wmat \!\otimes\! \Hmat$, where $\mbox{vec}(\cdot)$ denotes the vectorization operation, $\otimes$ the Kronecker product. The large measurement matrix $\Phimat \!\in\!\mathbb{R}^{hw\!\times\!HW}$ is replaced by two small matrices $\Hmat \!\in\!\mathbb{R}^{ h\!\times\!H}$, $\Wmat \!\in\!\mathbb{R}^{ w\!\times\!W}$ to reconstruct an image $\Xmat \!\in\!\mathbb{R}^{H\!\times\!W}$ from the measurements $\Ymat \!\in\!\mathbb{R}^{h\!\times\!w}$ with a compressive ratio $\eta \!=\!{hw \mathord{\left/
 {\vphantom {hw HW}} \right.
 \kern-\nulldelimiterspace} HW}$.
The inverse problem is therefore formulated as
\begin{equation}
\label{eq:kron}
\setlength{\abovedisplayskip}{0.1cm}
\setlength{\belowdisplayskip}{0.1cm}
 {\widehat \Xmat} \!=\! \underset{\Xmat}{\arg\min} f(\Xmat) \!+\! \lambda g({\Xmat}),
\end{equation}
where $f \!\!=\!\! \frac{1}{2} \! \left \| {\Ymat} \!-\! \Hmat \Xmat {\Wmat^{\top}} \right \|_{F}^{2}$, ${\left\| \cdot \right\|_{F}}$ denotes the Frobenius norm.
As a result, large matrix calculations on $\Phimat$ are equivalent to efficient calculations on $\Hmat$ and $\Wmat$, resulting in a significant decrease on space complexity from $\mathcal{O}(\eta H^2W^2)$ to $\mathcal{O}({\sqrt\eta}H^2\!+\!{\sqrt\eta}W^2)$ and time complexity from $\mathcal{O}(\eta H^2W^2)$ to $\mathcal{O}(\eta H^2W\!+\!{\sqrt\eta}HW^2)$ $(H \!\le \!W)$~\cite{wang2023saunet}.
Note that the physical masks in SPI cameras still correspond to the rows of $\Phimat \!=\! \Wmat \!\otimes\! \Hmat$.

\subsection{Proximal Algorithms: HQS and ADMM}
\label{subsec:algo}
HQS and ADMM are generally developed to solve \cref{eq:prob}. Next, we exploit both towards \cref{eq:kron}. By introducing an auxiliary variable $\Zmat \!\in\!\mathbb{R}^{H\!\times\!W}$, the unconstrained \cref{eq:kron} can be rewritten as a constrained optimization problem:
\begin{equation} \label{eq:split}
\setlength{\abovedisplayskip}{0.1cm}
\setlength{\belowdisplayskip}{0.1cm}
 ({\widehat \Xmat},{\widehat \Zmat}) \!=\! \underset{\Xmat, \Zmat}{\arg\min} f(\Zmat) \!+\! \lambda  g({\Xmat}), ~~s.t.~ \Zmat\!=\!\Xmat.
\end{equation}
HQS handles the constraint of \cref{eq:split} as a penalty term by minimizing
\begin{equation} \label{eq:hqs}
\setlength{\abovedisplayskip}{0.1cm}
\setlength{\belowdisplayskip}{0.1cm}
\mathcal{L}_{\mu}(\Xmat, \Zmat) \!=\! f(\Zmat) \!+\! \lambda  g({\Xmat}) \!+\! \frac{\mu}{2} \Vert {\Zmat} \!-\! {\Xmat} \Vert_{F}^{2},
\end{equation}
where $\mu \!>\! 0$. The minimizer of \cref{eq:split} is the saddle point of $\mathcal{L}$, which can be found by alternative proximal operators as
\begin{equation}\label{eq:hqs1}
\setlength{\abovedisplayskip}{0.1cm}
\setlength{\belowdisplayskip}{0.1cm}
{\tt Prox}_{f}\!:~~{\Zmat}^{k\!+\!1} \!=\! \underset{\Zmat}{\arg\min} ~f(\Zmat) \!+\! \frac{\mu}{2} \Vert {\Zmat} \!-\! {\Xmat}^{k} \Vert_{F}^{2},~~~~~~
\end{equation}
\begin{equation}\label{eq:hqs2}
\setlength{\abovedisplayskip}{0.1cm}
\setlength{\belowdisplayskip}{0.1cm}
{\tt Prox}_{g}\!:~~{\Xmat}^{k\!+\!1} \!=\!  \underset{\Xmat}{\arg\min} ~\lambda g(\Xmat) \!+\! \frac{\mu}{2} \Vert {\Xmat} \!-\! {\Zmat}^{k\!+\!1} \Vert_{F}^{2},
\end{equation}
where $\Xmat^{0} \!\!=\!\! \Hmat^{\top} \Ymat \Wmat$. Instead of adding a simple penalty term in HQS, ADMM re-formulates \cref{eq:split} using the augmented Lagrangian as
\begin{equation} \label{eq:admm}
\setlength{\abovedisplayskip}{0.1cm}
\setlength{\belowdisplayskip}{0.1cm}
\mathcal{L}_{\mu}(\Xmat, \!\Zmat, \!\Umat) \!=\! f(\Zmat) \hspace{-1.3pt}+\hspace{-1.3pt} \lambda  g({\Xmat}) \hspace{-1.3pt}+\hspace{-1.3pt} \Umat\tsp\!(\Zmat \hspace{-1.3pt}-\hspace{-1.3pt} \Xmat) \hspace{-1.3pt}+\hspace{-1.3pt} \frac{\mu}{2} \Vert {\Zmat} \hspace{-1.3pt}-\hspace{-1.3pt} {\Xmat} \Vert_{F}^{2},
\end{equation}
where $\Umat$ is the Lagrange multiplier. \cref{eq:admm} can also be solved by alternative proximal operators as
\begin{equation}\label{eq:admm1}
\setlength{\abovedisplayskip}{0.1cm}
\setlength{\belowdisplayskip}{0.1cm}
{\tt Prox}_{f}\!:~ {\Zmat}^{k\!+\!1} \!=\! \underset{\Zmat}{\arg\min} ~f(\Zmat) \hspace{-1.2pt}+\hspace{-1.2pt} \frac{\mu}{2} \Vert {\Zmat} \hspace{-1.2pt} -\hspace{-1.2pt} ({\Xmat}^{k} \hspace{-1.2pt}-\hspace{-1.2pt} \frac{1}{\mu}{\Umat}^{k}) \Vert_{F}^{2},
\end{equation}
\begin{equation}\label{eq:admm2}
\setlength{\abovedisplayskip}{0.1cm}
\setlength{\belowdisplayskip}{0.1cm}
~{\tt Prox}_{g}\!: {\Xmat}^{k\!+\!1} \!=\! \underset{\Xmat}{\arg\min} ~\lambda g(\Xmat) \!+\! \frac{\mu}{2} \Vert {\Xmat} \!-\! ({\Zmat}^{k\!+\!1} \!+\! \frac{1}{\mu}{\Umat}^{k}) \Vert_{F}^{2},
\end{equation}
with ${\Umat}^{k\!+\!1} \!=\! {\Umat}^{k} \!+\! \mu ({\Zmat}^{k\!+\!1} \!-\! {\Xmat}^{k\!+\!1})$.

In SPI paradigm, $f$ is a differentiable convex function and $\Phimat$ is generally set to be row-orthogonal, \ie, $\Wmat\Wmat\tsp \!\otimes\! \Hmat\Hmat\tsp \!=\! \Imat$. In this case, ${\tt Prox}_{f}$ has a closed form:
\begin{equation}\label{eq:hqs11}
\setlength{\abovedisplayskip}{0.1cm}
\setlength{\belowdisplayskip}{0.1cm}
{\tt Prox}_{f}(\Pmat) \!=\!  \Pmat \!+\! \frac{1}{{1 \!+\! \mu}} {\Hmat\tsp(\Ymat \!\!-\! \!\Hmat \Pmat \Wmat\tsp)\Wmat},
\end{equation}
where $\Pmat \!=\! \Xmat^{k}$ or $({\Xmat}^{k} \!-\! \frac{1}{\mu}{\Umat}^{k})$ for HSQ or ADMM.
\cref{eq:hqs11} is also a gradient descent step.
With a proper regularizer $g$,
HQS and ADMM could converge to stationary points of \cref{eq:kron} by iterating
\begin{equation}\label{eq:prox_algo}
\setlength{\abovedisplayskip}{0.1cm}
\setlength{\belowdisplayskip}{0.1cm}
{\Xmat}^{k\!+\!1} \!=\! {\tt Prox}_{g} \circ {\tt Prox}_{f}({\Xmat}^{k}).
\end{equation}

\subsection{Plug-and-Play}
PnP approaches typically replace ${\tt Prox}_{g}$ in \cref{eq:prox_algo} with an off-the-shelf deep denoiser~\cite{zhang2017beyond,zhang2018ffdnet,zhang2021plug} $\mathcal{D}_{\theta}$ with learned weights $\theta$, namely iterating
\begin{equation}\label{eq:pnp}
\setlength{\abovedisplayskip}{0.1cm}
\setlength{\belowdisplayskip}{0.1cm}
\Xmat^{k\!+\!1} \!=\!\mathcal{D}_{\theta} \circ {\tt Prox}_{f} (\Xmat^{k}).
\end{equation}
On the upside, $\mathcal{D}_{\theta}$ serves as a generic image prior to reduce unknown artifacts (including structural noises introduced by ${\tt Prox}_{f}$) regardless of the imaging tasks and the degree of degradation, namely high generalization. For SPI cameras, the generalization make PnP algorithms flexible for varying CRs.
Besides, $\mathcal{D}_{\theta}$ can be interpreted as a proximal operator to guarantee the convergence under non-expansiveness assumptions or Lipschitz constraints~\cite{ryu2019plug,hurault2022gradient,hurault2022proximal,fangs,hu2024stochastic}.
On the downside, PnP approaches are typically limited in reconstruction accuracy and speed since pretrained $\mathcal{D}_{\theta}$ lacks task-specific knowledge and a large number of iterations and cumbersome parameter adjustment are necessary.

\begin{figure*}[th!]
\centering
  \vspace{-4mm}
\includegraphics[width=0.9\linewidth]{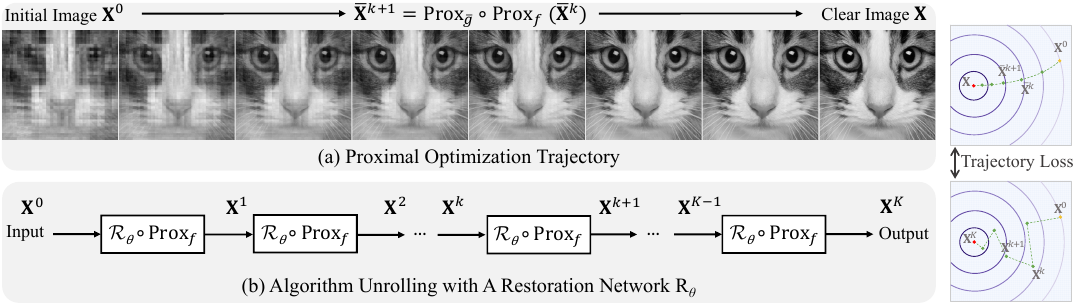}
\vspace{-2mm}
\caption{Proximal algorithm unrolling via trajectory loss.}
\vspace{-3mm}
\label{fig:proxunroll}
\end{figure*}

\subsection{Unrolling}
Unrolling approaches first replace ${\tt Prox}_{g}$ in \cref{eq:prox_algo} with a lightweight neural network $\mathcal{N}$, \ie, $\Xmat^{k\!+\!1} \!=\!\mathcal{N} \circ {\tt Prox}_{f} (\Xmat^{k})$, then unroll it with few iterations, finally train it end-to-end with degraded-clean image pairs produced by a specific degradation model $\Ymat \!\!=\!\! \Hmat \Xmat {\Wmat\tsp\!} \!\!+\!\! \Emat$, therefore formulated as 
\begin{equation}\label{eq:unroll}
\setlength{\abovedisplayskip}{0.1cm}
\setlength{\belowdisplayskip}{0.1cm}
\Xmat^{K} \!=\! \left[ \mathop \bigcirc\limits_{k\!=\!0}^{K\!-\!1}  \mathcal{N}_{\theta^{k}} \circ {\tt Prox}_{f} \right] (\Xmat^{0}),
\end{equation}
where $\Xmat^{0}\!=\! \Hmat^{\top} \Ymat \Wmat$ and $\Xmat^{K}$ denote the input and the output respectively, and $\bigcirc$ is the compound composition operator.
Instead of using a pretrained $\mathcal{D}_{\theta}$ in PnP algorithms, unrolling networks learn a small set of subnets $\{\mathcal{N}_{\theta^{k}}\}_{k=0}^{K\!-\!1}$ with a same structure but different weights, thereby belonging to multi-stage neural networks.
Although unrolling networks resemble proximal algorithms that are widely known to be convergent, end-to-end training could only ensure that the final result $\Xmat^{K}$ is close to the ground truth $\Xmat$ but cannot guarantee that the intermediate results $\{\Xmat^{k}\}_{k=1}^{K\!-\!1}$ are improved step-by-step, thereby lacking the convergence guarantee~\cite{10741959} (see~\cref{fig:iterate}).
After degradation-specific training, unrolling networks usually produce SOTA results with fast inference (due to very few iterations), but need be fine-tuned or even retrained when the degradation (\eg, CR in SPI) changes.
Consequently, performance is achieved at the cost of generalization.
How to make unrolling networks flexible for varying CRs is a key challenge in SPI.


\section{Proximal Algorithm Unrolling}
In this section, we first introduce a proximal algorithm unrolling approach with interpretability and convergence guarantees, named ProxUnroll, via a proximal trajectory (PT) loss function.
Subsequently, we propose an efficient CNN-Transformer architecture to realize ProxUnroll.

\subsection{Proximal Unrolling via Trajectory Loss} 
\label{subsec:method1}
The main differences between PnP in \cref{eq:pnp} and unrolling in \cref{eq:unroll} involve that, $(i)$ the iteration number is often large in PnP but small in unrolling; $(ii)$ one pretrained deep denoiser $\mathcal{D}_{\theta}$ is reused across iterations in PnP but $K$ subnets $\{\mathcal{N}_{\theta^{k}}\}_{k=0}^{K\!-\!1}$ are end-to-end optimized for $K$ iterations in unrolling; $(iii)$ PnP $\mathcal{D}_{\theta}$ could loosely correspond to ${\tt Prox}_{g}$ but what unrolled $\mathcal{N}_{{\theta}^{k}}$ has learned is unclear.
By integrating the characteristics of PnP and unrolling, we define a proximal unrolling (ProxUnroll) framework as
\begin{equation}\label{eq:proxunroll}
\setlength{\abovedisplayskip}{0.1cm}
\setlength{\belowdisplayskip}{0.1cm}
\Xmat^{K} \!=\! \left[ \mathop \bigcirc\limits_{k\!=\!0}^{K\!-\!1}  \mathcal{R}_{\theta} \circ {\tt Prox}_{f} \right] (\Xmat^{0}), ~~s.t.~~ \mathcal{R}_{\theta} \!\to\! {\tt Prox}_{g},
\end{equation}
where $\mathcal{R}_{\theta}$ is a neural network, which can be expressed as the proximal operator of a regularizer $g$.
In \cref{eq:proxunroll}, the remaining challenge is to satisfy the condition $\mathcal{R}_{\theta} \!\to\! {\tt Prox}_{g}$.

Essentially, ${\tt Prox}_{g}$ is an input-to-output movement towards the minimum of a regularizer $g$ such that the output is between the input and the minimum.
Given a degraded image as input, the output image is bound to be restored if the minimum is the corresponding clear image.
In this case, ${\tt Prox}_{g}$ is an image restoration operator.
Motivated by this point, we define the squared Euclidean distance between a degraded image $\Xmat'$ and its clear image $\Xmat$ as an explicit regularization function:
\begin{equation}\label{eq:regu}
\setlength{\abovedisplayskip}{0.1cm}
\setlength{\belowdisplayskip}{0.1cm}
{\bar g} (\Xmat') \!=\! \frac{1}{2} \Vert {\Xmat'} \!-\! {\Xmat} \Vert_{F}^{2}.
\end{equation}
By inserting ~\cref{eq:regu} into ~\cref{eq:hqs2,eq:admm2}, ${\tt Prox}_{\bar g}$ has the following closed form:
\begin{equation}\label{eq:prox_gt1}
\setlength{\abovedisplayskip}{0.1cm}
\setlength{\belowdisplayskip}{0.1cm}
{\tt Prox}_{\bar g} (\Qmat) \!= \! \frac{\mu {\Qmat} + \lambda {\Xmat}}{\mu + \lambda},
\end{equation}
where $\Qmat \!=\! \Zmat^{k\!+\!1}$ or $({\Zmat}^{k\!+\!1} \!+\! \frac{1}{\mu}{\Umat}^{k})$ for HQS or ADMM, $\mu$ and $\lambda$ control the strength of being close to $\Qmat$ and ${\Xmat}$ respectively.
By combining \cref{eq:hqs11,eq:prox_gt1}, an ideal explicit proximal algorithm is
\begin{equation}\label{eq:prox_gt}
\setlength{\abovedisplayskip}{0.1cm}
\setlength{\belowdisplayskip}{0.1cm}
\Xmat^{k\!+\!1} \!=\! {\tt Prox}_{\bar g} \circ {\tt Prox}_{f}(\Xmat^{k}),
\end{equation}
where $\Xmat^{k} \!\to\! \Xmat$ if $k \!\to\! \infty$.
The convergence rate is closely related to $\mu$ and $\lambda$.
If ${\mu \mathord{\left/
 {\vphantom {\mu  \lambda}} \right.
 \kern-\nulldelimiterspace} \lambda} \!=\! 0$, ${\tt Prox}_{\bar g}$ executes an one-step restoration from an arbitrarily degraded image to its clear image.
${\tt Prox}_{\bar g}$ is an explicit image restorer conditioned on the ground truth.
During supervised training, we use ${\tt Prox}_{\bar g}$ to guide the optimization of $\mathcal{R}_{\theta}$ in \cref{eq:proxunroll} such that $\mathcal{R}_{\theta}$ serves as an implicit image restorer to approximate ${\tt Prox}_{\bar g}$.

\begin{figure*}[th!]
\centering
\vspace{-4mm}
\includegraphics[width=0.91\linewidth]{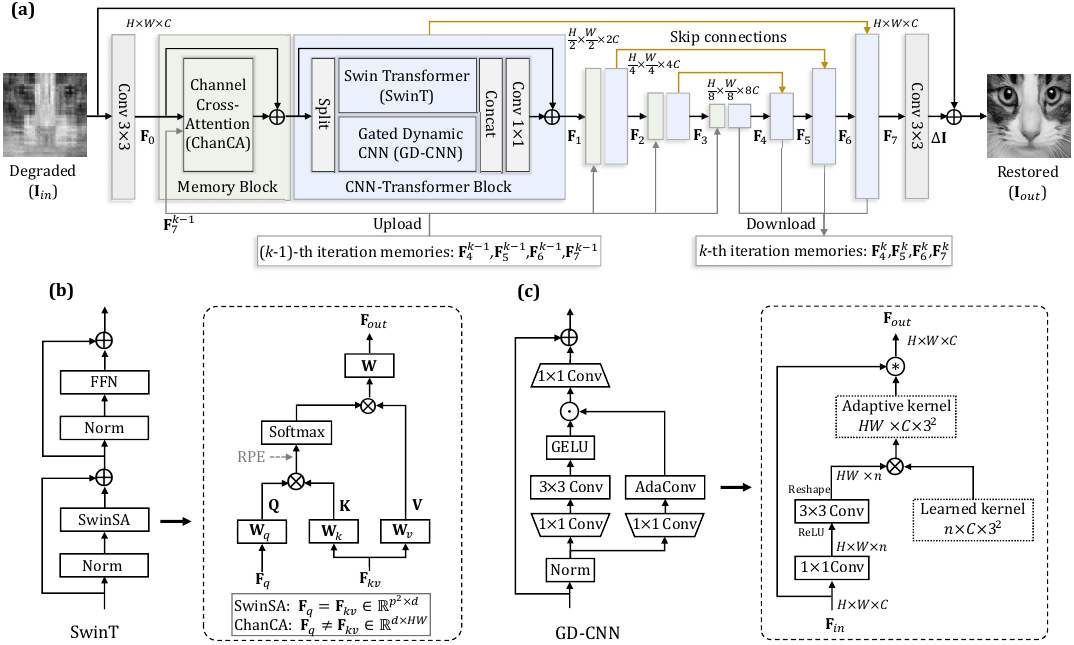}
\vspace{-1mm}
\caption{Illustration of the proposed deep image restorer (DIR) $\mathcal{R}_{\theta}$.}
\vspace{-4mm}
\label{fig:net}
\end{figure*}

As shown in \cref{fig:proxunroll}, a perfect proximal optimization trajectory $\Xmat^{0} \!\to\! \bar\Xmat^{1}  \!\to\! \cdots \!\to\! {\bar\Xmat^{K\!-\!1}} \!\to\! \Xmat$ can be produced from every $(\Xmat^0,\Xmat)$ training pair via \cref{eq:prox_gt} to supervise the end-to-end training of algorithm unrolling in~\cref{eq:proxunroll}.
The proximal iterations ${\tt Prox}_{\bar g} \circ {\tt Prox}_{f}$ are guaranteed with fast convergence towards the ground truth such that unrolled iterations $\mathcal{R}_{\theta} \circ {\tt Prox}_{f}$ are optimized to have the similar capability, namely $\mathcal{R}_{\theta} \!\to\! {\tt Prox}_{g}$.
To this end, we propose a proximal trajectory (PT) loss function as
\begin{equation}
\label{eq:loss}
\setlength{\abovedisplayskip}{0.1cm}
\setlength{\belowdisplayskip}{0.1cm}
\begin{array}{l}
{{\bf PL} = \mathop \sum\limits_{k\!=\!0}^{K\!-\!1} \alpha_{k}
\Vert \Xmat^{k\!+\!1} -  \bar\Xmat^{k\!+\!1} \Vert_{F}^{2}},\\
s.t.\left\{ {\begin{array}{*{20}{c}}
{\Xmat^{k\!+\!1} = \mathcal{R}_{\theta} \circ {\tt Prox}_{f}(\Xmat^{k})~~~~~}\\
{\bar\Xmat^{k\!+\!1} = {\tt Prox}_{\bar g} \circ {\tt Prox}_{f}(\bar\Xmat^{k})}
\end{array}} \right.
\end{array}
\end{equation}
where $\bar\Xmat^{0} \!=\! \Xmat^{0}$, $\bar\Xmat^{K} \!=\! \Xmat$, $\alpha_{k}$ controls the proximity strength of $k$-th iteration.
Trainable parameters involve ${\theta}, \{\mu^{k}\}_{k\!=\!0}^{K\!-\!1}$ in $\mathcal{R}_{\theta} \circ {\tt Prox}_{f}$ and $\{ {\bar\mu}^{k}, {\bar\lambda}^{k}  \}_{k\!=\!0}^{K\!-\!2}$ in ${\tt Prox}_{\bar g} \circ {\tt Prox}_{f}$.
For $\bar\Xmat^{K} \!=\! \Xmat$, $({\bar\mu}^{K\!-\!1}, {\bar\lambda}^{K\!-\!1}) $ is set to be $(0,1)$.
PT loss function is designed to train any unrolled network $\mathcal{R}_{\theta }$ to approximate the proximal operator ${\tt Prox}_{\bar g}$, thereby with interpretability.
Since ${\tt Prox}_{\bar g}$ is an explicit image restorer, $\mathcal{R}_{\theta }$ functions as a deep image restorer (DIR).




\subsection{Deep Image Restorer} 
\label{subsec:method2}
To unlock the potential of ProxUnroll, we propose an efficient CNN-Transformer architecture as DIR. 

\noindent{\bf Overall Pipeline.}
As shown in \cref{fig:net}~(a), DIR is a 4-level asymmetric encoder-decoder architecture.
Given a degraded image $\Imat_{in} \!\in\! \mathbb{R}^{H\!\times\!W\!\times\!C_{in}}$, one $3 \!\times\! 3$ convolution layer is first applied to extract shallow features $\Fmat_{0} \!\in\! \mathbb{R}^{H\!\times\!W\!\times\!C}$.
Next, $\Fmat_{0}$ are passed through $3$ encoder stages, each of which contains one memory block, a stack of hybrid CNN-Transformer blocks, and one downsampling layer.
From level-$1$ to level-$3$, the goal of the encoder is to progressively reduce the spatial resolution by half while
doubling channel capacity, thereby yielding multi-scale low-level features $\{\Fmat_{1},\Fmat_{2},\Fmat_{3}\}$. For example, the encoder produces $\Fmat_{3}\!\in\! \mathbb{R}^{ \frac{H}{8} \!\times\!\frac{W}{8}\!\times\!8C}$ at level-$3$.
At level-$4$, a bottleneck stage is between the encoder and the decoder to produce $\Fmat_{4}$. 
The decoder is similar to the encoder except for having no memory block and replacing the downsampling layer by one upsampling layer.
Form level-$3$ to level-$1$, the decoder progressively
produces high-level features $\{\Fmat_{5},\Fmat_{6},\Fmat_{7}\}$ while recovering the spatial resolution  and channel capability.
Finally, one $3 \!\times\! 3$ convolution layer transforms the features $\Fmat_{7}$ into the residual image $\Delta\Imat \!\in\! \mathbb{R}^{H\!\times\!W\!\times\!C_{in}}$.
The restored image is therefore obtained by $ \Imat_{out} = \Imat_{in} + \Delta \Imat$.

In these downsampling and upsampling layers, one $2\!\times\! 2$ strided convolution and one $2\!\times\! 2$ transposed convolution are employed respectively.
Skip connections are built between the encoder and the decoder for efficient information propagation.
Memory block and hybrid CNN-Transformer block are the core components. Next, we introduce them in detail.

\noindent{\bf Memory Block.}
As mentioned previously, typical unrolling networks in~\cref{eq:unroll} cannot guarantee the progressive refinement of intermediate results~\cite{10741959} (see \cref{fig:iterate}) due to end-to-end training.
That is also partially caused by that informative representations cannot be propagated across iterations.
Hence, the artifacts in intermediate results may be the manifestations of informative representations in the image domain.
As analyzed in the supplementary material, we attempt to reduce or remove the intermediate artifacts of SAUNet~\cite{wang2023saunet}, and
the resulting outputs become worse.

To address this problem, we propose a memory block (MB) to propagate informative representations in a side path.
MB is composed of one channel cross-attention (ChanCA), similar to existing multi-Dconv head transposed attention (MDTA)~\cite{zamir2022restormer}.
As shown in \cref{fig:net}~(a), we directly consider the output features of the bottleneck and decoder at iteration $k$ as current memories $\{\Fmat_{4}^{k},\Fmat_{5}^{k},\Fmat_{6}^{k},\Fmat_{7}^{k}\}$.
By producing queries from $\{\Fmat_{4}^{k},\Fmat_{5}^{k},\Fmat_{6}^{k},\Fmat_{7}^{k}\}$ and keys/values from $\{\Fmat_{4}^{k\!-\!1},\Fmat_{5}^{k\!-\!1},\Fmat_{6}^{k\!-\!1},\Fmat_{7}^{k}\}$, MB adaptively aggregates previous useful representations into current iteration for efficient information propagation by ChanCA, formulated as
\begin{equation}\label{eq:chanca}
\setlength{\abovedisplayskip}{0.1cm}
\setlength{\belowdisplayskip}{0.1cm}
\begin{aligned}
{\Fmat\!_{out}} \!=\! {\tt ChanCA}(\Fmat\!_{q}, \Fmat\!_{kv}) \!=\! \Wmat\!_{p} \!\left[ {\tt softmat}(\Qmat  \Kmat\!\tsp) \Vmat \right],\\
\Qmat \!=\! \Wmat_{p}^{q}  \Fmat_{q} \Wmat_{d}^{q}, \Kmat \!=\!  \Wmat_{p}^{k}  \Fmat_{kv} \! \Wmat_{d}^{k}, \Vmat \!=\! \Wmat_{p}^{v}  \Fmat_{kv} \! \Wmat_{d}^{v},
\end{aligned}
\end{equation}
where $\Fmat_{q}, \Fmat_{kv} \!\in\! \mathbb{R}^{d\!\times\!HW} $ denote the heads of current and previous memory features respectively, $\Qmat$ and $\{\Kmat,\Vmat\}$ are produced from $\Fmat_{q}$ and $\Fmat_{kv}$ by a point-wise convolution $\Wmat_{p}^{(\cdot)}$ followed by a $3\times3$ depth-wise convolution $\Wmat_{d}^{(\cdot)}$.
At the first iteration, MB can automatically be invalid due to the lack of available memories, \ie $\Fmat\!_{out}\!=\!0$ if $\Fmat\!_{kv}\!=\!0$.
Such design leads to two advantages: $i)$ cross-iteration information propagation mechanism is built regardless iterations;
$ii)$ informative representations are adaptively activated.

\noindent{\bf CNN-Transformer Block.}
We propose a hybrid CNN-Transformer block (CTB) as the basic processing unit. The core idea of CTB is to integrate high- and low-frequency modeling ability of CNN  and Transformer~\cite{park2022vision} while keeping high dependency for input.
As shown in \cref{fig:net}~(a), an input feature is firstly split into two parts evenly, then Swin Transformer (SwinT) and gated dynamic CNN (GD-CNN) process them in parallel, finally two parts are fused by channel concatenation and a point-wise convolution.

As depicted in \cref{fig:net}~(b), SwinT is mainly powered by a shifted-window self-attention (SwinSA) and a feed-forward network (FFN) composed of a MLP with a $3\times3$ depth-wise convolution inserted.
SwinSA can be formulated as
\begin{equation}\label{eq:swin}
\setlength{\abovedisplayskip}{0.1cm}
\setlength{\belowdisplayskip}{0.1cm}
\begin{aligned}
\Fmat\!_{out} \!=\! {\tt SwinSA}(\Fmat\!_{in}) \!=\! \left[ {\tt softmat} (\Qmat  \Kmat\!\tsp \!\!+\! \Bmat) \Vmat \right] \! \Wmat,\\
\Qmat \!=\! \Fmat\!_{in} \Wmat^{q}, \Kmat \!=\!  \Fmat\!_{in}  \Wmat^{k}, \Vmat \!=\! \Fmat\!_{in} \Wmat^{v},~~~~~
\end{aligned}
\end{equation}
where $\Fmat\!_{in} \!\in\! \mathbb{R}^{p^{2}\!\times\!d}$ is the head at a $p \times p$ window, $\Wmat^{(\cdot)}  \!\in\! \mathbb{R}^{d\!\times\!d}$ denotes a learnable linear projection matrix, and $\Bmat$ denote a learnable relative position encoding (RPE).
As opposed to ChanCA in \cref{eq:chanca}, SwinSA performs self-attention mechanism on spatial dimensions.

As depicted in \cref{fig:net}~(c), GD-CNN is a gated structure of normal convolutions (Convs) and an adaptive convolution (AdaConv). 
AdaConv is a simple implementation of transforming a group of static (learned) convolution kernels into dynamic one varying with the input, formulated as 
\begin{equation}\label{eq:swin}
\setlength{\abovedisplayskip}{0.1cm}
\setlength{\belowdisplayskip}{0.1cm}
\begin{aligned}
\Fmat \!=\! {\tt Reshape}({\tt Conv}_{3\!\times\!3}({\tt ReLU}({\tt Conv}_{1\!\times\!1}(\Fmat\!_{in})))),\\
\Fmat\!_{out} \!=\! (\Fmat \Wmat ) \circledast  \Fmat\!_{in},~~~~~~~~~~~~~~~~~~~~~
\end{aligned}
\end{equation}
where $\Fmat\!_{in}, \Fmat\!_{out} \!\in\! \mathbb{R}^{H\!\times\!W\times\!C}$ denote the input and the output, $\Fmat  \!\in\! \mathbb{R}^{HW\!\times n}$ the input-dependent kernel coefficients, $\Wmat \!\in\! \mathbb{R}^{n\!\times\!C\!\times\!3^2}$ the kernel weights of $n$ $3\times3$ depth-wise convolutions, and $\circledast$ performs the convolution operator.
Similar to attention mechanism, AdaConv is also input-dependent.
Differently, AdaConv updates convolution weights from a group of learned kernel weights while attention produces attention weights by the similarity between features.



\section{Experiments}


The proposed DIR $\mathcal{R}_{\theta}$ in~\cref{subsec:method2} is plugged in HQS- or ADMM-unrolling framework and then is end-to-end optimized using the proposed PL loss function in~\cref{subsec:method1}, leading to two proximal unrolling networks: {HQS-ProxUnroll} and {ADMM-ProxUnroll}.
HQS/ADMM-ProxUnroll is composed of $6$ iterations of $\mathcal{R}_{\theta} \circ {\tt Prox}_{f}$.
Next, we conduct experiments to estimate them.
We adopt $400$ images from BSD500~\cite{5557884} to generate $20,000$ training samples using data augmentation techniques following~\cite{shi2019image,shi2019scalable,song2021memory,zhang2020amp,shen2022transcs,mou2022deep,song2023optimization,wang2023saunet,qu2024dual}.
Similar to the training of PnP image denoisers~\cite{zhang2017beyond,zhang2018ffdnet,zhang2021plug}, the compressive ratio varies within $\left[0.01,0.50\right]$ and the resolution varies from $256\!\times \!256$ pixels to $512\!\times \!512$ pixels to cover as wider degradations as possible.
The initial learning rate is ${1\times10^{-3}}$ and gradually reduced to ${1\times10^{-4}}$.
We estimate {HQS-ProxUnroll} and {ADMM-ProxUnroll} on Set11 and BSD68 benchmark datasets and real captured SPI data.
Regarding CBSD68, all experiments are performed on the luminance channel of YCbCr space, similar to~\cite{shi2019image,shi2019scalable,zhang2020optimization,zhang2020amp,song2021memory,mou2022deep,song2023optimization}.
Following~\cite{wang2023saunet,qu2024dual}, the measurement (\ie, degradation) matrices ($\Hmat,\Wmat$) are set to be floating-point and trainable for simulation experiments and are the reordered Hadamard matrices for real experiments.
More experiment results are in the supplementary material.

\begin{figure*}[t]
\centering
\vspace{-4mm}
\includegraphics[width=0.97\linewidth]{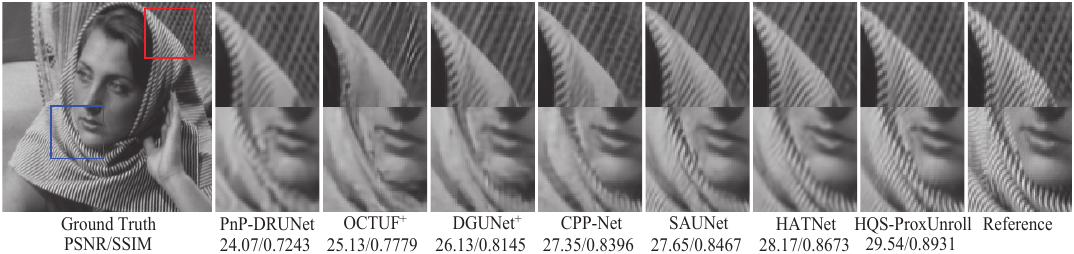}
\vspace{-2mm}
\caption{Visual comparisons on ``barbara'' from Set11 dataset at CR~$=0.04$.} 
\vspace{-1mm}
\label{fig:grayresult}
\end{figure*}

\begin{figure*}[t]
\centering
\vspace{-1mm}
\includegraphics[width=0.97\linewidth]{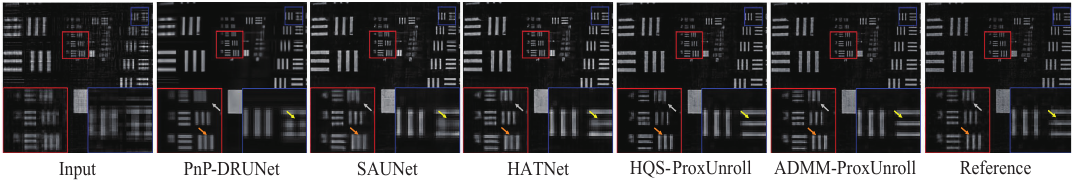}
\vspace{-2mm}
\caption{Visual comparisons on real SPI data ``resolution target'' reported in~\cite{qu2024dual} at CR~$=0.15$.} 
\vspace{-2mm}
\label{fig:realresult1}
\end{figure*}

\begin{table*}[!htbp]
\centering
\caption{Average PSNR/SSIM of different methods on Set11 and CBSD68 datasets with different compressive ratios. The best and second best results are \textbf{highlighted} and \underline{underlined}.
Blank cells (-) means that there is not an available model at the compressive ratio.
}
\vspace{-2mm}
\label{tab:result}
\setlength{\tabcolsep}{5pt}
\centering
\scalebox{0.85}{
\begin{tabular}{|m{-5cm}|m{-5cm}|m{-5cm}|m{-5cm}|m{-5cm}|m{-5cm}|m{-5cm}|m{-5cm}|m{-5cm}|}

\toprule[1.25pt]
\multicolumn{1}{c|}{\multirow{2}{*}{Dataset}}  &
\multicolumn{1}{c|}{\multirow{2}{*}{Method}}   & 
\multicolumn{1}{c|}{\multirow{2}{*}{Flexibility}}  &
\multicolumn{5}{c}{Compressive Ratio (CR)}  &
 \multicolumn{1}{|c}{\multirow{2}{*}{Average}} 
\\ 
\cline{4-8} 
\multicolumn{1}{c|}{} &  
\multicolumn{1}{c|}{}  &
\multicolumn{1}{c|}{}  &
\multicolumn{1}{c}{0.01}  & 
\multicolumn{1}{c}{0.04}  & 
\multicolumn{1}{c}{0.10}  & 
\multicolumn{1}{c}{0.25}  & 
\multicolumn{1}{c}{0.50} &
\multicolumn{1}{|c}{}  
 \\
\midrule
\multicolumn{1}{c|}{}  &
\multicolumn{1}{l|}{ISTA-Net$^+$~\cite{zhang2018ista}}  &
\multicolumn{1}{l|}{}  &
\multicolumn{1}{c}{17.42/0.4029} &
\multicolumn{1}{c}{21.32/0.6037} & 
\multicolumn{1}{c}{26.64/0.8087} & 
\multicolumn{1}{c}{32.59/0.9254} & 
\multicolumn{1}{c}{38.11/0.9707} & 
\multicolumn{1}{|c}{27.22/0.7423}         \\
\multicolumn{1}{c|}{}  &
\multicolumn{1}{l|}{CSNet$^+$~\cite{shi2019image}}   & 
\multicolumn{1}{l|}{}  &
\multicolumn{1}{c}{20.67/0.5411} &
\multicolumn{1}{c}{24.83/0.7480} & 
\multicolumn{1}{c}{28.34/0.8580} & 
\multicolumn{1}{c}{33.34/0.9387} & 
\multicolumn{1}{c}{38.47/0.9796} & 
\multicolumn{1}{|c}{29.13/0.8131}          \\
\multicolumn{1}{c|}{}  &
\multicolumn{1}{l|}{OPINE-Net$^+$~\cite{zhang2020optimization}} &
\multicolumn{1}{l|}{}  &
\multicolumn{1}{c}{20.15/0.5340} &
\multicolumn{1}{c}{25.69/0.7920} &
\multicolumn{1}{c}{29.81/0.8884} & 
\multicolumn{1}{c}{34.86/0.9509} & 
\multicolumn{1}{c}{40.17/0.9797} & 
\multicolumn{1}{|c}{32.63/0.9397}          \\
\multicolumn{1}{c|}{}  &
\multicolumn{1}{l|}{AMP-Net~\cite{zhang2020amp}} &
\multicolumn{1}{l|}{}  &
\multicolumn{1}{c}{20.21/0.5429} &
\multicolumn{1}{c}{25.27/0.7821} &
\multicolumn{1}{c}{29.43/0.8880} & 
\multicolumn{1}{c}{34.71/0.9532} & 
\multicolumn{1}{c}{40.66/0.9827} & 
\multicolumn{1}{|c}{30.06/0.9679}          \\
\multicolumn{1}{c|}{}  & 
\multicolumn{1}{l|}{TransCS~\cite{shen2022transcs}}   &
\multicolumn{1}{l|}{}  &
\multicolumn{1}{c}{20.22/0.5431} &
\multicolumn{1}{c}{25.41/0.7883	} & 
\multicolumn{1}{c}{29.54/0.8877} & 
\multicolumn{1}{c}{35.06/0.9548} & 
\multicolumn{1}{c}{40.49/0.9815} & 
\multicolumn{1}{|c}{29.42/0.8311}          \\
\multicolumn{1}{c|}{}  &
\multicolumn{1}{l|}{CSformer~\cite{ye2023csformer}}   & 
\multicolumn{1}{c|}{Single}  &
\multicolumn{1}{c}{21.58/0.6075} &
\multicolumn{1}{c}{26.28/0.8062} & 
\multicolumn{1}{c}{29.79/0.8883} & 
\multicolumn{1}{c}{34.81/0.9527} & 
\multicolumn{1}{c}{40.73/0.9824} & 
\multicolumn{1}{|c}{30.64/0.8474}          \\

\multicolumn{1}{c|}{}  &
\multicolumn{1}{l|}{MADUN~\cite{song2021memory}} &
\multicolumn{1}{c|}{}  &
\multicolumn{1}{c}{20.28/0.5572} &
\multicolumn{1}{c}{25.71/0.8042} &
\multicolumn{1}{c}{30.20/0.9016} & 
\multicolumn{1}{c}{35.76/0.9601} & 
\multicolumn{1}{c}{41.00/\underline{0.9837}} & 
\multicolumn{1}{|c}{30.59/0.8373}          \\
\multicolumn{1}{c|}{Set11}  &
\multicolumn{1}{l|}{DGUNet$^+$~\cite{mou2022deep}} &
\multicolumn{1}{c|}{}  &
\multicolumn{1}{c}{{22.15/0.6113}} &
\multicolumn{1}{c}{{26.82/0.8230}} &
\multicolumn{1}{c}{{30.92/0.9088}} & 
\multicolumn{1}{c}{{36.18/0.9616}} & 
\multicolumn{1}{c}{41.24/\underline{0.9837}} & 
\multicolumn{1}{|c}{31.46/0.8577}          \\
\multicolumn{1}{c|}{}  &
\multicolumn{1}{l|}{OCTUF$^+$~\cite{song2023optimization}} &
\multicolumn{1}{l|}{}  &
\multicolumn{1}{c}{21.94/0.5989} &
\multicolumn{1}{c}{26.54/0.8150} &
\multicolumn{1}{c}{30.73/0.9036} & 
\multicolumn{1}{c}{36.10/0.9607} & 
\multicolumn{1}{c}{41.35/\textbf{0.9838}} & 
\multicolumn{1}{|c}{31.33/0.8524}          \\
\multicolumn{1}{c|}{}  &
\multicolumn{1}{l|}{CPP-Net~\cite{guo2024cpp}}      & 
\multicolumn{1}{l|}{}  &
\multicolumn{1}{c}{{22.19/0.6135}} &
\multicolumn{1}{c}{{27.23/0.8337}} & 
\multicolumn{1}{c}{{31.27/0.9135}} & 
\multicolumn{1}{c}{{36.35/0.9631}} &
\multicolumn{1}{c}{{-}} & 
\multicolumn{1}{|c}{-}          \\
\multicolumn{1}{c|}{}  &
\multicolumn{1}{l|}{SAUNet~\cite{wang2023saunet}}      & 
\multicolumn{1}{l|}{}  &
\multicolumn{1}{c}{{22.43/0.6134}} &
\multicolumn{1}{c}{{27.80/0.8353}} & 
\multicolumn{1}{c}{{32.15/0.9147}} & 
\multicolumn{1}{c}{{37.11/0.9628}} &
\multicolumn{1}{c}{41.91/\textbf{0.9838}} & 
\multicolumn{1}{|c}{32.28/0.8620}          \\
\multicolumn{1}{c|}{}  &
\multicolumn{1}{l|}{HATNet~\cite{qu2024dual}}      & 
\multicolumn{1}{l|}{}  &
\multicolumn{1}{c}{{\underline{22.54}/0.6162}} &
\multicolumn{1}{c}{{27.98/0.8382}} & 
\multicolumn{1}{c}{{32.26/0.9182}} & 
\multicolumn{1}{c}{{37.24/0.9634}} &
\multicolumn{1}{c}{\textbf{42.05}/\textbf{0.9838}} & 
\multicolumn{1}{|c}{32.41/0.8640}          \\
\cmidrule{2-9}
\multicolumn{1}{c|}{}  &
\multicolumn{1}{l|}{COAST~\cite{you2021coast}}      & 
\multicolumn{1}{c|}{Multiple}  &
\multicolumn{1}{c}{{10.45/0.2144$^*$}} &
\multicolumn{1}{c}{{22.90/0.6971$^*$}} & 
\multicolumn{1}{c}{{28.69/0.8618}} & 
\multicolumn{1}{c}{{33.90/0.9399$^*$}} &
\multicolumn{1}{c}{{38.94/0.9744}} & 
\multicolumn{1}{|c}{26.98/0.7375}          \\

\cmidrule{2-9}
\multicolumn{1}{c|}{}  &
\multicolumn{1}{l|}{PnP-DnCNN~\cite{ryu2019plug}}      & 
\multicolumn{1}{l|}{}  &
\multicolumn{1}{c}{{21.68/0.5682}} &
\multicolumn{1}{c}{{26.45/0.7969}} & 
\multicolumn{1}{c}{{29.57/0.8712}} & 
\multicolumn{1}{c}{{34.24/0.9402}} &
\multicolumn{1}{c}{{39.78/0.9735}} & 
\multicolumn{1}{|c}{30.34/0.8300}          \\

\multicolumn{1}{c|}{}  &
\multicolumn{1}{l|}{PnP-DRUNet~\cite{zhang2021plug}}      & 
\multicolumn{1}{l|}{}  &
\multicolumn{1}{c}{{21.75/0.5738}} &
\multicolumn{1}{c}{{26.81/0.8074}} & 
\multicolumn{1}{c}{{30.16/0.8861}} & 
\multicolumn{1}{c}{{35.00/0.9476}} &
\multicolumn{1}{c}{{40.54/0.9767}} & 
\multicolumn{1}{|c}{30.85/0.8383}          \\
\multicolumn{1}{c|}{}  &
\multicolumn{1}{l|}{HQS-ProxUnroll}      & 
\multicolumn{1}{c|}{Arbitrary}  &
\multicolumn{1}{c}{\textbf{22.76}/\underline{0.6300}} &
\multicolumn{1}{c}{\underline{28.23}/\underline{0.8450}} & 
\multicolumn{1}{c}{\underline{32.51}/\underline{0.9225}} & 
\multicolumn{1}{c}{\underline{37.33}/\underline{0.9635}} &
\multicolumn{1}{c}{41.92/\underline{0.9837}} & 
\multicolumn{1}{|c}{\underline{32.55}/\underline{0.8689}}          \\
\multicolumn{1}{c|}{}  &
\multicolumn{1}{l|}{ADMM-ProxUnroll}      & 
\multicolumn{1}{c|}{}  &
\multicolumn{1}{c}{\textbf{22.76}/\textbf{0.6307}} &
\multicolumn{1}{c}{\textbf{28.30}/\textbf{0.8452}} & 
\multicolumn{1}{c}{\textbf{32.55}/\textbf{0.9226}} & 
\multicolumn{1}{c}{\textbf{37.35}/\textbf{0.9639}} &
\multicolumn{1}{c}{\underline{41.97}/\textbf{0.9838}} & 
\multicolumn{1}{|c}{\textbf{32.59}/\textbf{0.8692}}          \\

\midrule[1pt]

\multicolumn{1}{c|}{}  &
\multicolumn{1}{l|}{ISTA-Net$^+$~\cite{zhang2018ista}} & 
\multicolumn{1}{l|}{}  &
\multicolumn{1}{c}{19.14/0.4158} &
\multicolumn{1}{c}{22.17/0.5486} & 
\multicolumn{1}{c}{25.32/0.7022} & 
\multicolumn{1}{c}{29.36/0.8525} & 
\multicolumn{1}{c}{34.04/0.9424} & 
\multicolumn{1}{|c}{26.01/0.6923}          \\
\multicolumn{1}{c|}{}  &
\multicolumn{1}{l|}{CSNet$^+$~\cite{shi2019image}}      & 
\multicolumn{1}{l|}{}  &
\multicolumn{1}{c}{22.21/0.5100} &
\multicolumn{1}{c}{25.43/0.6706} &
\multicolumn{1}{c}{27.91/0.7938} & 
\multicolumn{1}{c}{31.12/0.9060} &
\multicolumn{1}{c}{36.76/0.9638} & 
\multicolumn{1}{|c}{28.69/0.7688}          \\
\multicolumn{1}{c|}{}  &
\multicolumn{1}{l|}{OPINE-Net$^+$~\cite{zhang2020optimization}} &
\multicolumn{1}{l|}{}  &
\multicolumn{1}{c}{22.11/0.5140} &
\multicolumn{1}{c}{25.20/0.6825} &
\multicolumn{1}{c}{27.82/0.8045} & 
\multicolumn{1}{c}{31.51/0.9061} & 
\multicolumn{1}{c}{36.35/0.9660} & 
\multicolumn{1}{|c}{28.60/0.7746}          \\
\multicolumn{1}{c|}{}  &
\multicolumn{1}{l|}{AMP-Net~\cite{zhang2020amp}} &
\multicolumn{1}{l|}{}  &
\multicolumn{1}{c}{22.28/0.5319} &
\multicolumn{1}{c}{25.27/0.6930} &
\multicolumn{1}{c}{27.88/0.8118} & 
\multicolumn{1}{c}{31.80/0.9154} & 
\multicolumn{1}{c}{37.03/0.9722} & 
\multicolumn{1}{|c}{28.85/0.7849}          \\
\multicolumn{1}{c|}{}  & 
\multicolumn{1}{l|}{TransCS~\cite{shen2022transcs}}   &
\multicolumn{1}{l|}{}  &
\multicolumn{1}{c}{22.28/0.5318} &
\multicolumn{1}{c}{25.28/0.6881} & 
\multicolumn{1}{c}{27.86/0.8086} & 
\multicolumn{1}{c}{31.74/0.9121	} & 
\multicolumn{1}{c}{36.81/0.9699} & 
\multicolumn{1}{|c}{28.79/0.7821}          \\
\multicolumn{1}{c|}{}  &
\multicolumn{1}{l|}{CSformer~\cite{ye2023csformer}}   &
\multicolumn{1}{c|}{Single}  &
\multicolumn{1}{c}{22.81/0.5566} &
\multicolumn{1}{c}{25.73/0.6956} & 
\multicolumn{1}{c}{28.05/0.8045} & 
\multicolumn{1}{c}{31.82/0.9106	} & 
\multicolumn{1}{c}{37.14/0.9766} & 
\multicolumn{1}{|c}{29.11/0.7889}          \\
\multicolumn{1}{c|}{}  &
\multicolumn{1}{l|}{MADUN~\cite{song2021memory}} &
\multicolumn{1}{c|}{}  &
\multicolumn{1}{c}{22.08/0.5247} &
\multicolumn{1}{c}{25.36/0.6985} &
\multicolumn{1}{c}{28.18/0.8219} & 
\multicolumn{1}{c}{{32.27/0.9219}} & 
\multicolumn{1}{c}{37.23/0.9733} & 
\multicolumn{1}{|c}{29.02/0.7880}          \\
\multicolumn{1}{c|}{}  &
\multicolumn{1}{l|}{DGUNet$^+$~\cite{mou2022deep}} &
\multicolumn{1}{c|}{}  &
\multicolumn{1}{c}{22.13/0.5215} &
\multicolumn{1}{c}{25.45/0.6986} &
\multicolumn{1}{c}{28.13/0.8165} & 
\multicolumn{1}{c}{31.97/0.9158} & 
\multicolumn{1}{c}{37.04/0.9718} & 
\multicolumn{1}{|c}{28.94/0.7848}          \\
\multicolumn{1}{c|}{BSD68}  &
\multicolumn{1}{l|}{OCTUF$^+$~\cite{song2023optimization}} &
\multicolumn{1}{c|}{}  &
\multicolumn{1}{c}{{22.78/0.5413}} &
\multicolumn{1}{c}{{25.65/0.6999}} &
\multicolumn{1}{c}{{28.28/0.8177}} & 
\multicolumn{1}{c}{32.24/0.9185} & 
\multicolumn{1}{c}{{37.41/0.9729}} & 
\multicolumn{1}{|c}{29.27/0.7900}          \\
\multicolumn{1}{c|}{}  &
\multicolumn{1}{l|}{CPP-Net~\cite{guo2024cpp}}      & 
\multicolumn{1}{c|}{}  &
\multicolumn{1}{c}{{22.95/0.5475}} &
\multicolumn{1}{c}{{25.81/0.7068}} & 
\multicolumn{1}{c}{{28.41/0.8227}} & 
\multicolumn{1}{c}{{32.25/0.9188}} &
\multicolumn{1}{c}{{-}} & 
\multicolumn{1}{|c}{-}          \\

\multicolumn{1}{c|}{}  &
\multicolumn{1}{l|}{SAUNet~\cite{wang2023saunet}}      & 
\multicolumn{1}{c|}{}  &
\multicolumn{1}{c}{{23.11/0.5460}} &
\multicolumn{1}{c}{{26.23/0.7050}} & \multicolumn{1}{c}{{29.25/0.8251}} & 
\multicolumn{1}{c}{{33.67/0.9243}} & 
\multicolumn{1}{c}{\textbf{39.28}/0.9751} & 
\multicolumn{1}{|c}{30.31/0.7951}          \\
\cmidrule{2-9}
\multicolumn{1}{c|}{}  &
\multicolumn{1}{l|}{COAST~\cite{you2021coast}}      & 
\multicolumn{1}{c|}{Multiple}  &
\multicolumn{1}{c}{{12.08/0.2584$^*$}} &
\multicolumn{1}{c}{{22.77/0.5953$^*$}} & 
\multicolumn{1}{c}{{26.28/0.7422}} & 
\multicolumn{1}{c}{{30.07/0.8703$^*$}} &
\multicolumn{1}{c}{{34.74/0.9497}} & 
\multicolumn{1}{|c}{25.19/0.6832}          \\

\cmidrule{2-9}
\multicolumn{1}{c|}{}  &
\multicolumn{1}{l|}{PnP-DnCNN~\cite{ryu2019plug}}      & 
\multicolumn{1}{l|}{}  &
\multicolumn{1}{c}{{22.27/0.5221}} &
\multicolumn{1}{c}{{24.85/0.6327}} & 
\multicolumn{1}{c}{{27.22/0.7658}} & 
\multicolumn{1}{c}{{30.95/0.8762}} &
\multicolumn{1}{c}{{36.47/0.9601}} & 
\multicolumn{1}{|c}{28.35/0.7514}          \\
\multicolumn{1}{c|}{}  &
\multicolumn{1}{l|}{PnP-DRUNet~\cite{zhang2021plug}}      & 
\multicolumn{1}{l|}{}  &
\multicolumn{1}{c}{{22.81/0.5308}} &
\multicolumn{1}{c}{{25.30/0.6670}} & 
\multicolumn{1}{c}{{27.79/0.7852}} & 
\multicolumn{1}{c}{{31.38/0.8931}} &
\multicolumn{1}{c}{{36.86/0.9624}} & 
\multicolumn{1}{|c}{28.83/0.7677}          \\
\multicolumn{1}{c|}{}  &
\multicolumn{1}{l|}{HQS-ProxUnroll}      & 
\multicolumn{1}{c|}{Arbitrary}  &
\multicolumn{1}{c}{\underline{23.51}/\underline{0.5645}} &
\multicolumn{1}{c}{\underline{26.52}/\underline{0.7194}} & 
\multicolumn{1}{c}{\underline{29.42}/\underline{0.8332}} & 
\multicolumn{1}{c}{\underline{33.76}/\underline{0.9282}} &
\multicolumn{1}{c}{{39.22}/\underline{0.9760}} & 
\multicolumn{1}{|c}{\underline{30.49}/\underline{0.8043}}          \\
\multicolumn{1}{c|}{}  &
\multicolumn{1}{l|}{ADMM-ProxUnroll}      & 
\multicolumn{1}{l|}{}  &
\multicolumn{1}{c}{\textbf{23.53}/\textbf{0.5649}} &
\multicolumn{1}{c}{\textbf{26.54}/\textbf{0.7200}} &
\multicolumn{1}{c}{\textbf{29.43}/\textbf{0.8334}} &
\multicolumn{1}{c}{\textbf{33.77}/\textbf{0.9284}} &
\multicolumn{1}{c}{\underline{39.23}/\textbf{0.9761}} &
\multicolumn{1}{|c}{\textbf{30.50}/\textbf{0.8046}}          \\
\bottomrule[1.25pt]
\end{tabular}
}
\vspace{-3mm}
\end{table*}

\vspace{-1mm}
\subsection{Results on Benchmark Datasets}
\vspace{-1mm}
SPI reconstruction approaches are closely related to a great number of image compressed sensing approaches.
The proposed HQS-ProxUnroll and ADMM-ProxUnroll are compared with previous approaches on benchmark datasets at five widely-used CRs $\{0.01,0.04,0.10,0.25,0.50\}$.
Comparative approaches involve a single-stage network (CSNet$^+$~\cite{shi2019image}), unrolling (multi-stage) networks (ISTA-Net$^+$~\cite{zhang2018ista}, OPINENet$^+$~\cite{zhang2020optimization}, AMP-Net~\cite{zhang2020amp}, COAST~\cite{you2021coast}, TransCS~\cite{shen2022transcs}, MADUN~\cite{song2021memory}, DGUNet$^+$~\cite{mou2022deep}, CSformer~\cite{ye2023csformer}, OCTUF$^+$~\cite{song2023optimization}, SAUNet~\cite{wang2023saunet}, CPP-Net~\cite{guo2024cpp}, HATNet~\cite{qu2024dual}), and PnP algorithms (PnP-DnCNN~\cite{ryu2019plug} and PnP-DRUNet~\cite{zhang2021plug}).
\cref{tab:result} reports the reconstruction results of different approaches.
Towards reconstruction accuracy, the proposed HQS-ProxUnroll and ADMM-ProxUnroll are close and both of them outperform previous approaches on average.
As visualized in~\cref{fig:grayresult}, our HQS/ADMM-ProxUnroll can recover more image details.
Towards the flexibility for CR, most previous approaches are separately trained for a specific CR and thus cannot generalize to different CRs.
COAST, trained on multiple CRs $\{0.10,0.20,0.30,0.40,0.50\}$, is flexible only for seen CRs, not for unseen CRs (\eg $0.01$).
Our HQS/ADMM-ProxUnroll and PnP algorithms (PnP-DnCNN, PnP-DRUNet) can flexibly handle arbitrary CRs with a single model. 
Moreover, we compare HQS/ADMM-ProxUnroll with several competitive approaches in terms of multiply-accumulate operations (MACs), model parameters (Params), and runtime.
As reported in~\cref{tab:metric}, our HQS/ADMM-ProxUnroll achieves an all-around superiority over these approaches.
Overall, our HQS/ADMM-ProxUnroll achieve not only SOTA performance and efficiency but also the same flexibility as PnP algorithms.



\begin{table}[t]
\vspace{-4mm}
\caption{MACs (G), Params (M), and runtime (sec.) comparisons on $256\!\times\! 256$ images at CR~$=0.25$. PnP-DRUNet performs $50$ iterations for best performance.} 
\vspace{-3mm}
\label{tab:metric}
\setlength{\tabcolsep}{5pt}
\scalebox{0.75}{
\begin{tabular}{c|ccccc}
\toprule[1.25pt]
\multirow{2}{*}{Metric} & 
CPP-Net & SAUNet & HATNet & PnP-DRUNet & HQS/ADMM-   \\ 
&
\cite{guo2024cpp} & \cite{wang2023saunet} & \cite{qu2024dual} & \cite{zhang2021plug} & ProxUnroll  \\ 
\hline
\hline

\multirow{2}{*}{MACs} &  \multirow{2}{*}{ 153.47} &  \multirow{2}{*}{\underline{143.05}} & \multirow{2}{*}{494.42} & 7175.10 & \textbf{107.73} \\
& & & & (143.5$\times$50)  &  (17.95$\times$6)
\\ \hline
Params  &  12.31 &  \underline{10.53}  &  31.28 & 32.64 & \textbf{3.90} \\ \hline
Runtime &   0.41 &    \underline{0.35}  &  0.60  & 0.65 & \textbf{0.27} \\
\bottomrule[1.25pt]
\end{tabular} 
}
\vspace{-5mm}
\end{table}

\vspace{-1mm}
\subsection{Results on Real Captured Data}
\vspace{-1mm}
We estimate the proposed HQS/ADMM-ProxUnroll on real captured SPI data.
Among the networks mentioned previously, only SAUNet~\cite{wang2023saunet} and HATNet~\cite{qu2024dual} are practical for real SPI cameras. The remaining networks are developed under the assumption that images can be compressively sampled block by block, which is out of line with real SPI cameras.
Hence, we compare HQS/ADMM-ProxUnroll with SAUNet, HATNet, and PnP-DRUNet.
As shown in~\cref{fig:realresult1}, PnP-DRUNet can recovers well on low frequencies (\eg, shapes and edges) instead of high frequencies (\eg, textures and details), agreeing with that Gaussian denoisers could smooth out rather than remove unknown artifacts~\cite{zhang2021plug}.
HQS-ProxUnroll and ADMM-ProxUnroll are close and both of them have a better reconstruction ability than the competitors. 
In practice, SAUNet and HATNet need a separate model for a specific CR.
A single model of HQS-ProxUnroll or ADMM-ProxUnroll is flexible enough for varying CRs, similar to PnP algorithms.

\vspace{-1mm}
\subsection{Ablation Study}
\vspace{-1mm}
To offer an insight into ProxUnroll, we conduct ablation experiments on PL loss function (\cref{subsec:method1}) and DIR (\cref{subsec:method2}).
ProxUnroll uses PL loss function to train DIR as approximation of the proximal operator of an explicit restoration regularizer in \cref{{eq:regu}} to guarantee the fast convergence.

\begin{figure}[t]
\centering
\vspace{-4mm}
\includegraphics[width=0.75\linewidth]{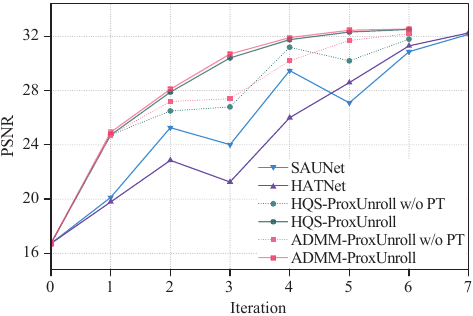}
\vspace{-3mm}
\caption{Convergence curves of different unrolling methods.
} 
\vspace{-1.5mm}
\label{fig:iterate}
\end{figure}

To demystify the influence of PL loss function, the convergence curves of different methods are depicted in \cref{fig:iterate}.
SAUNet and HATNet, both composed of $7$ iterations, reconstruct images with twists and turns since their MSE loss function only ensure a satisfactory finial result.
This implies that their intermediate subnets cannot be interpreted as the proximal restoration operators and function more like black-box neural networks.
Their lack of proximity also elucidates why prior unrolling networks require retraining when the degradation changes.
With PL-based training, our HQS/ADMM-ProxUnroll achieves fast and stable convergence. 
By replacing PL loss with regular MSE loss during training, their convergence curves also become unstable.

To demystify the influence of key designs in DIR, we conduct ablation experiments as shown in \cref{tab:ablation}.
The proposed DIR is mainly powered by three components: ChanCA in memory block, SwinT and GD-CNN in CNN-Transformer block.
ChanCA is used to adaptively propagate informative representations between adjacent iterations.
The combination of SwinT and GD-CNN is to integrate the high- and low-frequency modeling ability while keeping high dependency for input.
GD-CNN is powered by the proposed AdaConv. 
Clearly, ChanCA results in a 1.09 dB improvement in PSNR with a slight increase in MACs, parameters, and runtime.
By discarding SwinT and doubling the channel capability of GD-CNN, the entire network becomes a CNN, leading to a 0.26 dB decrease in PSNR and an all-round increase in other metrics.
Replacing AdaConv with one depth-wise convolution causes a 0.15 dB decrease in PSNR but has minimal impact on MACs, parameters, and runtime.
These results validate the effectiveness and efficiency of ChanCA, SwinT, and AdaConv.


\begin{table}[t]
\vspace{-2mm}
\caption{Ablation experiments for different components of
our deep image restorer (DIR) on Set11 dataset at CR~$=0.10$.} 
\vspace{-3mm}
\label{tab:ablation}
\setlength{\tabcolsep}{4pt}
\scalebox{0.835}{
\begin{tabular}{ccc|cccc}
\toprule[1.25pt]
ChanCA  & SwinT & AdaConv & PSNR & MACs & Params & Runtime  \\ \hline\hline
     & \checkmark & \checkmark &  31.42 &  94.85 &   3.10 &   0.22     \\ 
\checkmark &     &\checkmark &  32.25 &  128.67 & 4.65  &   0.34     \\ 
\checkmark &\checkmark &         &  32.36 & 101.99 &  3.85 &  0.21 \\ 
\checkmark &\checkmark & \checkmark  &  32.51  & 107.73&  3.90 &  0.27   \\ 
\bottomrule[1.25pt]
\end{tabular}
}
\vspace{-5.5mm}
\end{table}

\vspace{-0.5mm}
\section{Conclusion}
\vspace{-0.5mm}
We propose ProxUnroll to train any unrolled network to approximate the proximal operator of an explicit restoration regularizer using a proximal trajectory (PL) loss function.
Besides, we propose an efficient CNN-Transformer architecture as deep image restorer (DIR) to arm ProxUnroll.
Extensive experiments demonstrate that our ProxUnroll achieves not only SOTA performance and efficiency but also the same flexibility as PnP algorithms. 
\section*{Acknowledgements}
This work was supported by National Key R\&D Program of China (2024YFF0505603), the National Natural Science Foundation of China (grant number 62271414), Zhejiang Provincial Distinguished Young Scientist Foundation (grant number LR23F010001), Zhejiang “Pioneer” and “Leading Goose” R\&D Program (grant number 2024SDXHDX0006, 2024C03182), the Key Project of Westlake Institute for Optoelectronics (grant number 2023GD007), the 2023 International Sci-tech Cooperation Projects under the purview of the “Innovation Yongjiang 2035” Key R\&D Program (grant number 2024Z126), Shanghai Municipal Science and Technology Major Project (2021SHZDZX0102), and the Fundamental Research Funds for the Central Universities.
{
    \small
    \bibliographystyle{ieeenat_fullname}
    \bibliography{main}
}


\end{document}